\documentclass[preprint,superscriptaddress, showpacs,showkeys,preprintnumbers,prb]{revtex4}

\usepackage{graphicx}

\usepackage{pslatex}
\usepackage[T1]{fontenc}
\usepackage{textcomp}

\newcommand{\sto}{SrTiO$_3$}
\newcommand{\bto}{BaTiO$_3$}
\newcommand{\cto}{CaTiO$_3$}
\newcommand{\lao}{LaAlO$_3$}

\begin{document}


\title{X-ray scattering from surfaces: discrete and continuous components of roughness}

\author{Darren Dale}
 \email{dd55@cornell.edu}
 \altaffiliation{Present Address: Cornell High Energy Synchrotron Source, Cornell University, Ithaca, New York 14853, USA}
 \affiliation{Department of Materials Science and Engineering, Cornell University, Ithaca, New York 14853, USA}
 \affiliation{Cornell Center for Materials Research, Cornell University, Ithaca, New York 14853, USA}
\author{Aaron Fleet}
\altaffiliation{Present Address: MIT Lincoln Laboratory, 244 Wood Street, Lexington, MA 02420, USA}
 \affiliation{School of Applied and Engineering Physics, Cornell University, Ithaca, New York 14853, USA}
 \affiliation{Cornell Center for Materials Research, Cornell University, Ithaca, New York 14853, USA}
\author{Y. Suzuki}
\affiliation{Department of Materials Science and Engineering, UC Berkeley, Berkeley, California 94720, USA}
\author{J. D. Brock}
 \affiliation{School of Applied and Engineering Physics, Cornell University, Ithaca, New York 14853, USA}
 \affiliation{Cornell Center for Materials Research, Cornell University, Ithaca, New York 14853, USA}

\date{\today}

\begin{abstract}
Incoherent surface scattering yields a statistical description of the surface, due to the ensemble averaging over many independently sampled volumes. Depending on the state of the surface and direction of the scattering vector relative to the surface normal, the height distribution is discrete, continuous, or a combination of the two. We present a treatment for the influence of multimodal surface height distributions on Crystal Truncation Rod scattering.  The effects of a multimodal height distribution are especially evident during \textit{in-situ} monitoring of layer-by-layer thin-film growth via Pulsed Laser Deposition.  We model the total height distribution as a convolution of discrete and continuous components, resulting in a broadly applicable parameterization of surface roughness which can be applied to other scattering probes, such as electrons and neutrons. Convolution of such distributions could potentially be applied to interface or chemical scattering. Here we find that this analysis describes accurately our experimental studies of $\langle 001 \rangle$ \sto\ annealing and homoepitaxial growth.
\end{abstract}

\maketitle

\section{Introduction}

The interaction of a scattering probe with a smooth, crystalline surface gives rise to streaks of intensity in reciprocal space known as Crystal Truncation Rods (CTRs). CTR intensity is sensitive to atomic-scale surface roughness such as the oscillatory surface roughness that arises during the so-called layer-by-layer~\cite{LCA_Stoop_1973_TSF_v17_p291} or polynuclear growth~\cite{FC_Frank_1974_JCG_v22_p233, GH_Gilmer_1980_JCG_v49_p465, W_van_Saarloos_1986_PRB_v33_p4927}. Reflection High Energy Electron Diffraction (RHEED) is one example of a technique that is often used for \textit{in-situ} monitoring of film growth via CTR intensity.~\cite{T_Terashima_1990_PRL_v65_p2684, H_Karl_1992_PRL_v69_p2939, G_Koster_1999_APL_v74_p3729, A_Ohtomo_2002_Nature_v419_p378, A_Ohtomo_2004_Nature_v427_p423} Recently, the large monochromatic flux available from synchrotron sources ($\approx10^{13}$ photons/second) have enabled time-resolved \textit{in-situ} X-ray scattering studies of film growth as well.~\cite{E_Vlieg_1988_PRL_v61_p2241, PH_Fuoss_1992_PRL_v69_p2791, DW_Kisker_1996_JCG_v163_p54, RL_Headrick_1996_PRB_v54_p14686, AR_Woll_1999_PRL_v83_p4349, MV_RamanaMurty_1999_PRB_v60_p16956, G_Eres_2002_APL_v80_p3379, A_Fleet_2005_PRL_v94_p036102, A_Fleet_2006_PRL_v96_p055508, PR_Willmott_2006_PRL_v96_p176102, JZ_Tischler_2006_PRL_v96_p226104} These \textit{in-situ} scattering techniques enable materials engineering at the atomic level via the carefully controlled deposition of sequences of single atomic layers.~\cite{K_Iijima_1992_JAP_v72_p2840, T_Tsurumi_1994_JJAP_v33_p5192, H_Tabata_1994_APL_v65_p1970, R_Takahashi_2002_ASS_v197_p532}

Growth of complex oxide thin films with atomic-scale control via scattering-based \textit{in-situ} monitoring has recently drawn intense interest. This is due in part to the broad range of materials properties that are manifest with changes in stoichiometry, even within a single structure family.  Pulsed Laser Deposition (PLD) is an attractive growth technique for such epitaxial complex oxide films and heterostructures, due to its ability to transfer complex stoichiometries from the target to the film.   Many interesting phenomena have recently been reported, including metallic behavior at the interface between two band insulators in \sto/\lao\ heterostructures~\cite{A_Ohtomo_2004_Nature_v427_p423}, superconductivity in BaCuO$_2$/SrCuO$_2$ heterostructures~\cite{G_Koster_2001_PC_v353_p167} (neither BaCuO$_2$ nor SrCuO$_2$ are superconducting in bulk), ferroelectricity in SrZrO$_3$/\sto\ heterostructures~\cite{T_Tsurumi_2004_APL_v85_p5016} (neither SrZrO$_3$ nor \sto\ is ferroelectric in bulk), and enhanced ferroelectric polarization in \cto/\sto/\bto\ heterostructures~\cite{HN_Lee_2005_Nature_v433_p395}.

Despite the increasing popularity of scattering probes as a method for investigating growth, the problem of directly comparing models of growth kinetics to scattering data remains a challenge. Such a comparison requires descriptions of the surface and the resulting scattered intensity that are complex enough to capture the physics of the problem, yet simple enough to lend themselves to, for example, least squares fitting. The influence of surface roughness on X-ray intensity can be readily calculated for Gaussian roughness~\cite{B_Vidal_1984_AO_v23_p1794} or discrete roughness on the order of the out-of-plane lattice spacing~\cite{CS_Lent_1984_SS_v139_p121} such as arises during nearly ideal layer-by-layer growth, but we require a more generally applicable model to analyze the evolution between these two limits. Previous reports have modeled continuous or discrete roughness independently of one another,~\cite{HN_Yang_1993} but real crystal surfaces have a miscut, and therefore have concomitant discrete and continuous components of surface roughness. 

Here, we treat the overall surface roughness as a convolution of discrete binomial and continuous Gaussian distributions, which results in a simple parameterization that is broadly applicable. We show that the X-ray intensity along the CTR associated with reciprocal lattice point $\mathbf{G}$ can be written in a simple, closed-form solution as
$$
\frac{I^{\mathbf{G}}(\mathbf{Q})}{I^{\mathbf{G}}_0(\mathbf{Q})} = \left[ 1-4p(1-p)\sin^2\left( \frac{Q_\perp c^\prime}{2} \right) \right]^n e^{ -\sigma_c^2 (Q_\perp-G_\perp)^2 },
$$
where $c^\prime$ is the discrete step height, $n$ and $p$ are the usual parameters in the binomial distribution, and $\sigma_c$ is the continuous RMS roughness. We present the application of this model to our studies of \textit{in-situ} X-ray scattering during annealing of $\langle 001 \rangle$ \sto\ and subsequent homoepitaxial growth via PLD, and find that our data is accurately described by a simple model for the dependence of $\sigma_c$, $n$ and $p$ with film thickness.

\section{Theory}

In this report, we present a closed form solution of the kinematic scattering theory for surfaces with roughness that is continuous, discrete, or a combination of the two. We begin by assuming that the first Born approximation is valid, i.e. scattering geometries that do not coincide with Bragg conditions. The scattered intensity can be written in the form $I(\mathbf{Q})=C(\mathbf{Q})|A(\mathbf{Q})|^2$, where $C(\mathbf{Q})$ is a prefactor associated with the scattering geometry, polarization, volume of the unit cell, and scattering cross-section.~\cite{D_Gibbs_1988_PRB_v38_p7303, ED_Specht_1993_JAC_v26_p166, Warren_1990} Assuming no surface reconstruction, the scattered intensity for a miscut crystal can be calculated by explicitly performing the kinematic sums:
\begin{equation}
\label{eq:crystal-sum}
I(\mathbf{Q}) = \sum_N C(\mathbf{Q}) \left| \sum_{\{\mathbf{s}_\parallel\}} \frac{F_{sub}(\mathbf{Q}) e^{-i\mathbf{Q}\cdot\mathbf{R}(\mathbf{s}_\parallel)} } {1-e^{iQ_zc_0}} \right|^2,
\end{equation}
where $\hbar \mathbf{Q}$ is the momentum transfer, $F_{sub}(\mathbf{Q})$ is the structure factor of the substrate, $c_0$ is the out-of-plane lattice parameter, $\mathbf{s}_\parallel$ is a vector in the plane of the average surface, $\mathbf{R}(\mathbf{s}_\parallel)$ is the position of the surface, $\sum_{\{\mathbf{s}_\parallel\}}$ is a sum over surface positions in each coherently illuminated region, and $\sum_N$ is a sum over coherent regions. X-ray scattering measurements can be classified as either coherent, partially coherent, or incoherent, depending on the size and coherence volume of the X-ray beam. Coherent scattering probes a single realization of the surface ($N=1$), while incoherent scattering probes the statistical behavior of an ensemble of realizations. This report will focus on incoherent scattering measurements. A simulated reciprocal-space map of the incoherent scattering intensity for an $\langle 001 \rangle$-oriented SrTiO$_3$ crystal with $0.198^\circ$ miscut is presented in figure \ref{fig:STO-rsmap}-(a). When the terrace size is small relative to the transverse coherence length, each reciprocal lattice point $\mathbf{G}$ has an associated CTR oriented parallel to the surface normal of the average surface. Figure \ref{fig:STO-rsmap}-(b) compares experimental $K$-scan data ($H=0$, $L=0.5$) with the model. The dominant peaks are associated with the $\langle 000 \rangle$ and $\langle 001 \rangle$ CTRs, and satellites can be observed which correspond to the $\langle 00\bar{1} \rangle$ and $\langle 002 \rangle$ rods. Note also the weak diffuse scattering, which is associated with irregularities in the terrace size and thermal diffuse scattering.

While such brute-force calculations are illuminating, they are quite computationally intensive.  Alternatively, statistical approaches can be used. Assuming translational invariance, it has previously been shown that the scattering intensity of a CTR associated with reciprocal lattice point $\mathbf{G}$ can in general be written for any surface height distribution~\cite{SK_Sinha_1988_PRB_v38_p2297, G_Held_1995_PRB_v51_p7262}
\begin{equation}
\label{eq:crystal-sum-statistical-general}
I_{total} \propto \sum_{\{\mathbf{s}_\parallel\}} \sum_{\{\mathbf{s}_\parallel^\prime\}} e^{-i (\mathbf{Q}-\mathbf{G}) \cdot (\mathbf{s}_\parallel - \mathbf{s}_\parallel^\prime)} \left\langle e^{-i (Q_\perp-G_\perp) [h(\mathbf{s}_\parallel)- h(\mathbf{s}_\parallel^\prime)]} \right\rangle,
\end{equation}
where $\mathbf{s}_\parallel - \mathbf{s}_\parallel^\prime$ is the in-plane vector between scatterers, $ Q_\perp-G_\perp = (\mathbf{Q}-\mathbf{G} ) \cdot \hat{\mathbf{n}}$, $h=\mathbf{r}\cdot\hat{\mathbf{n}}$ is the surface height as measured along the surface normal (as opposed to the terrace normal), and $\left\langle e^{-i (Q_\perp-G_\perp) [h(\mathbf{s}_\parallel) - h(\mathbf{s}_\parallel^\prime)]} \right\rangle$ is the height difference function.

Previous theoretical investigations, primarily for two-level systems such as ideal layer-by-layer growth on crystals without miscut, have shown that the total scattering intensity can be split into the sum of CTR and diffuse components $I_{total} = I_{CTR} + I_{diffuse}$.~\cite{SR_Andrews_1985_JPC_v18_p6427, CS_Lent_1984_SS_v139_p121, JM_Pimbley_1984_JVSTA_v2_p457, SK_Sinha_1996_PA_v231_p99}  This statistical interpretation of scattering has enabled extensive studies of thin-film growth.~\cite{HN_Yang_1993}


Here we focus on the CTR scattering intensity associated with reciprocal lattice point $\mathbf{G}$, which may be either specular or off-specular. Given $\Delta h(\mathbf{s}_\parallel) = h(\mathbf{s}_\parallel) - \langle h \rangle$, we assume delta-correlated height fluctuations $\langle \Delta h(\mathbf{s}_\parallel) \Delta h(\mathbf{s}_\parallel^\prime) \rangle = \sigma^2 \delta(\mathbf{s}_\parallel - \mathbf{s}_\parallel^\prime)$ such that
\begin{equation}
\label{eq:crystal-sum-statistical}
I^{\mathbf{G}}(\mathbf{Q}) = I_0^{\mathbf{G}}(\mathbf{Q}) \left| \left\langle e^{-i (Q_\perp-G_\perp) h(\mathbf{s}_\parallel)} \right\rangle \right|^2.
\end{equation}
For a Gaussian surface height distribution, the scattering intensity can be simplified to yield the familiar result
\begin{equation}
I^{\mathbf{G}}(\mathbf{Q}) = I_0^{\mathbf{G}}(\mathbf{Q}) e^{-(Q_\perp-G_\perp)^2 \sigma^2 }.
\label{eq:gaussian-roughness}
\end{equation}
For a flat crystal without miscut, the ideal intensity is~\cite{G_Held_1995_PRB_v51_p7262}
\begin{equation}
I_0^{\mathbf{G}}(\mathbf{Q}) = C(\mathbf{Q})\frac{\left|F_{sub}(\mathbf{Q})\right|^2 \delta^{(2)}(\mathbf{Q}_\parallel-\mathbf{G}_\parallel)} {\sin^2\left( \frac{Q_\perp c_0}{2} \right)}.
\end{equation}
For a crystal with a miscut that yields a surface normal along $\hat{\mathbf{n}}$, the ideal intensity (which neglects the surface roughness associated with the miscut itself), is
\begin{equation}
I_0^{\mathbf{G}}(\mathbf{Q}) = C(\mathbf{Q})\frac{\left|F_{sub}(\mathbf{Q})\right|^2 \delta^{(2)}[(\mathbf{Q}-\mathbf{G}) \times \hat{\mathbf{n}}]}{(Q_\perp-G_\perp)^2 }.
\end{equation}

An Atomic Force Microscopy (AFM) image of a $\langle 001 \rangle$ \sto\ surface is shown in figure \ref{fig:STO-miscut}-(a).  The existence of a miscut, with irregular terrace sizes, produces a Gaussian height distribution, and the resulting X-ray intensity can therefore be described by equation \ref{eq:gaussian-roughness}. Many surfaces, however, have more complicated height distributions. PLD, for example, results in the nucleation, growth, and coalescence of many small, two-dimensional islands on the surface. Previous \textit{in-situ} studies of PLD growth have shown that material that deposits on one of these small islands tends to diffuse to the step edge and incorporate into the underlying layer,~\cite{G_Eres_2002_APL_v80_p3379, A_Fleet_2005_PRL_v94_p036102, A_Fleet_2006_PRL_v96_p055508} enabling extended smooth growth. A simulation of the surface shown in figure \ref{fig:STO-miscut}-(a) after depositing one half monolayer via layer-by-layer growth is shown in figure \ref{fig:STO-miscut}-(b).  The system has evolved such that a bimodal distribution exists. Therefore, the surface height distribution has both continuous and discrete character. The discussion that follows is motivated by the need of a model that can handle the effect of such a surface height distribution on the scattered intensity.

In order to proceed, it is instructive to consider a special case: an ideally flat crystal with zero miscut and initial CTR intensity $I^{\mathbf{G}}_0(\mathbf{Q})$.  We then assume random, pulsed deposition where material does not deposit on the same site twice in a single pulse, and no surface diffusion occurs after adsorption. The surface height distribution is then given by the well-known binomial distribution: 
\begin{equation}
P_d(h) = \sum_k{n \choose k}p^k(1-p)^{n-k}\delta(h-kc^\prime),
\end{equation}
where $n$ is the number of pulses, $p$ is the fraction of the surface covered with new material during a single pulse, $c^\prime$ is the step height and $k$ is an index of overlayers. The mean is $\langle h\rangle=npc^\prime$ and the RMS surface roughness is $\sigma_d=c^\prime\sqrt{np(1-p)}$. The expectation value in equation \ref{eq:crystal-sum-statistical} can be simplified, using the binomial theorem:
\begin{eqnarray}
\left\langle e^{-i (Q_\perp-G_\perp) h(\mathbf{s}_\parallel)} \right\rangle & = & \int_{-\infty}^\infty P_d(h) e^{-i(Q_\perp-G_\perp) h}\mathrm{d}h \nonumber \\
& = & \sum_{k=0}^n {n \choose k} (p e^{-i(Q_\perp-G_\perp) c^\prime})^k (1-p)^{n-k} \nonumber \\
& = & \left( 1 - p + pe^{-i(Q_\perp-G_\perp) c^\prime} \right)^n
\label{eq:binomial-transform}.
\end{eqnarray}
The scattering intensity from a surface with a binomial height distribution is therefore
\begin{equation}
\frac{I^{\mathbf{G}}(\mathbf{Q})}{I^{\mathbf{G}}_0(\mathbf{Q})} = \left[ 1-4p(1-p)\sin^2\left( \frac{(Q_\perp-G_\perp) c^\prime}{2} \right) \right]^n
\label{eq:binomial-intensity}.
\end{equation}
Alternatively, this result can be derived by performing a sum of the overlayers, as described by Robinson,~\cite{IK_Robinson_1986_PRB_v33_p3830} and illustrated in appendix \ref{app:sum-overlayers}. There it is shown for the more general case where the amount of material that adsorbs during each pulse may vary, that the intensity after $n$ pulses is
\begin{equation}
\frac{I^{\mathbf{G}}(\mathbf{Q})}{I^{\mathbf{G}}_0(\mathbf{Q})} = \prod_n \left[ 1-4p_n(1-p_n)\sin^2\left( \frac{(Q_\perp-G_\perp) c^\prime}{2} \right) \right]
\label{eq:general-pulsed-intensity}.
\end{equation}
This more flexible description of the surface comes at the expense of additional parameters. In figure \ref{fig:compare-p}, three CTR intensity profiles for $\langle 001 \rangle$ \sto\ are simulated: an ideally flat substrate; a sample with $n=1$, $p=0.5$, $\sigma_d= 1.95$\AA; and a sample with $n=25$, $p=0.01$ and $\sigma_d= 1.94$\AA. The CTR intensity is most sensitive to the surface roughness in the so-called anti-Bragg scattering geometries, where $Q_\perp=\frac{m\pi}{c^\prime}, m=1,3,5,...$ Figure \ref{fig:compare-p} illustrates that the RMS surface roughness itself is not always sufficient to describe the surface statistically, since the two surfaces with nearly identical total roughness have very different CTR profiles. The $n,p$ parameterization provides the additional information necessary to model discrete surface roughness on the order of the out-of-plane lattice spacing.

Equation \ref{eq:binomial-intensity} can be rearranged, and by substituting $n=\sigma_d^2/{c^\prime}^2p(1-p)$, we arrive at an expression that describes how the scattered intensity varies with $\sigma_d,Q_z$ and $p$:
\begin{eqnarray}
\label{eq:I-vs-sigma-flat}
\frac{I^{\mathbf{G}}(\mathbf{Q})}{I_0^{\mathbf{G}}(\mathbf{Q})} = \exp \left\{ \frac{\sigma_d^2} {p(1-p){c^\prime}^2} \log_e \left[ 1 - 4 p \left( 1 - p \right) \sin^2 \left( \frac{(Q_\perp-G_\perp) c^\prime} {2} \right) \right] \right\}.
\end{eqnarray}
In the limit of $Q_\perp \rightarrow G_\perp$, equation \ref{eq:I-vs-sigma-flat} can be approximated using first order Taylor expansions, $\sin^2(x) \approx x^2/2$, $\log_e(1-x) \approx -x$. First expanding $sin^2(x)$, the expansion of the resulting log yields a simplified form of equation \ref{eq:I-vs-sigma-flat}, which is independent of $p$ and identical to the Gaussian roughness dependence in equation \ref{eq:gaussian-roughness}.

Alternatively, in the limit of $p\rightarrow0$, the log in equation \ref{eq:I-vs-sigma-flat} can be approximated regardless of the value of $Q_z$. In this limit, the binomial distribution exhibits Gaussian behavior, and the intensity can be written
\begin{equation}
\label{eq:p-app-0}
\lim_{p\rightarrow 0}\frac{I^{\mathbf{G}}(\mathbf{Q})}{I^{\mathbf{G}}_0(\mathbf{Q})} =  \exp \left[ \frac{ -4 \sigma_d^2} {{c^\prime}^2} \sin^2 \left( \frac{Q_\perp c^\prime} {2} \right) \right].
\end{equation}
Equation \ref{eq:p-app-0} is applicable to continuous deposition methods, as well as the limit where scattered intensity becomes insensitive to the discrete nature of the surface roughness. For experimental methods that are insensitive to the crystal miscut, equation \ref{eq:p-app-0} therefore also represents the appropriate model for continuous roughness, and is applicable anywhere along the CTR.

Having derived an expression for the influence of discrete roughness, we return our attention to the case where the surface height distribution is a continuous, multimodal function. The convolution of a binomial distribution defined by $n=1$ and $p=0.5$ with a Gaussian distribution will produce the surface height distribution shown in figure \ref{fig:STO-miscut}-(b).  Using the convolution theorem, we now have a solution for the effects of the total surface roughness on the CTR intensity.  The roughness has been parameterized into continuous and discrete contributions, which add in quadrature to yield the total surface roughness. For experimental methods that are insensitive to the crystal miscut, the total intensity including continuous and discrete contributions of roughness is written:
\begin{equation}
\label{eq:I-vs-sigma-general-int}
\frac{I(\mathbf{Q})}{I_0(\mathbf{Q})} =  \left[ 1-4p(1-p)\sin^2\left( \frac{Q_\perp c^\prime}{2} \right) \right]^n \exp \left[ \frac{ -4 \sigma_c^2} {c_0^2} \sin^2 \left( \frac{Q_\perp c^\prime} {2} \right) \right].
\end{equation}
For miscut crystals where the intensity of a single CTR is measured, the total intensity is
\begin{equation}
\label{eq:I-vs-sigma-general}
\frac{I^{\mathbf{G}}(\mathbf{Q})}{I^{\mathbf{G}}_0(\mathbf{Q})} = \left[ 1-4p(1-p)\sin^2\left( \frac{Q_\perp c^\prime}{2} \right) \right]^n e^{ -\sigma_c^2 (Q_\perp-G_\perp)^2 }.
\end{equation}

\section{Experimental Results}

We used surface-sensitive X-ray scattering to study the $\langle 001 \rangle$ \sto\ surface during annealing and homoepitaxy at the Cornell High Energy Synchrotron Source. Experiments were performed at the G3 hutch on the G-Line beamline. A double crystal synthetic multi-layer monochrometer with 1.5\% energy bandpass selected 10 KeV X-rays from the 48-pole wiggler spectrum, yielding $3\times10^{11}$ photons/s/mm$^2$ at the sample. The spot size on the sample was set with 2 mm vertical by 0.5 mm horizontal slit immediately before the sample and the detector resolution was set with a 2 mm vertical by 0.5 mm horizontal slit at 600 mm from the sample.

\subsection{Surface Roughening and Annealing}

In order to grow epitaxial films of complex oxides, it is necessary to elevate the substrate temperature in order to enhance diffusion and reorganization of the adsorbed species. We have found that the $\langle 001 \rangle$ \sto\ surface roughens during annealing to 300$^\circ$C. Figure \ref{fig:IvsL} shows experimental specular CTR intensity data from two $\langle 001 \rangle$ \sto\ samples under vacuum, one at room temperature (a), the other annealed at 300$^\circ$C (b). Both samples received a buffered HF etch treatment prior to measurement.~\cite{M_Kawasaki_1994_Science_v266_p1540} The HF etch yields an atomically smooth TiO$_2$-terminated \sto\ surface, with step edges that are one unit cell in height. During the heating process, we have observed a decline of CTR intensity in the $\langle 00\frac{1}{2} \rangle$ scattering geometry, which we attribute to an increased step edge density that may be related to a loss of oxygen from the \sto\ surface.~\cite{M_Lippmaa_1998_ASS_v130_p582} A least-squares fit for the room temperature data was performed using equation \ref{eq:p-app-0}, yielding $\sigma_c=1.52$\ \AA. The CTR intensity of the room temperature sample does not appear to be influenced by a discrete surface height distribution, as we would expect based on AFM results.
The CTR intensity of the 300$^\circ$C data, however, is very strongly influenced by discrete surface roughness; a least-squares fit was performed using equation \ref{eq:I-vs-sigma-general-int}, where $n=1$, $p=0.406$. The fit of the 300$^\circ$C data yields a continuous roughness $\sigma_c=0.715$ \AA\ and a discrete surface roughness of $\sigma_d=1.917$ \AA, yielding a total surface roughness of $\sigma=2.05$ \AA.

The temperature-dependent anti-Bragg scattering intensity for $\langle 001 \rangle$ \sto\ samples heated in $1\times10^{-6}$ Torr O$_2$ through 0.3 Torr O$_2$ is presented in figure \ref{fig:sto-roughnes-vs-T}. As the temperature increased, the scattering intensity initially decreased due to increased surface roughness. The temperature corresponding to minimum scattering intensity varied with O$_2$ pressure: $\approx250^\circ$C for 0.3 Torr, $\approx350^\circ$C for $1\times10^{-3}$ Torr, and $\approx600^\circ$C for $3\times10^{-6}$ Torr. As the temperature was increased futher, the scattering intensity increased as the surface roughness decreased. Increasing the temperature in 0.3 Torr O$_2$ resulted in improved surface roughness at lower temperatures. The baseline error of the roughness determination in figure \ref{fig:sto-roughnes-vs-T} is on the order of 10\%, which dominates the determination of the errorbars. These results indicate that the surface roughness of $\langle 001 \rangle$ \sto\ can be improved by annealing at elevated temperature, and that ramping the substrate temperature in an appropriate O$_2$ pressure results in improved surface quality at lower temperatures.

In figure \ref{fig:sto-roughnes-vs-time}, an $\langle 001 \rangle$ \sto\ substrate was heated to 800$^\circ$C in $1\times10^{-6}$ Torr O$_2$. The scattered anti-Bragg intensity was monitored as a function of time, showing modest improvement of the surface roughness. After one hour, the O$_2$ pressure was increased to $1\times10^{-3}$ Torr, resulting in a more rapid improvement of the scattered intensity and surface roughness. The total surface roughness in figures \ref{fig:sto-roughnes-vs-T} and \ref{fig:sto-roughnes-vs-time} does not approach zero, but a value corresponding to the roughness associated with the miscut. This result illustrates that at elevated temperatures, the use of an appropriate O$_2$ pressure causes the discrete roughness component to decrease at a faster rate, and that both continuous and discrete roughness should be considered during analysis.

\subsection{Epitaxial Growth}

Given an acceptable starting surface of $\langle 001 \rangle$ \sto, we can focus on homoepitaxial growth via PLD. A pulsed KrF excimer laser, with 248 nm wavelength and 30 ns pulse duration, was focused to provide a 200 MW/cm$^2$ power density at the surface of a single crystal \sto\ target. The repetition rate was 0.03 Hz, and the growth rate was 0.08 monolayer/pulse. The substrate surface temperature was 550$^\circ$C, and growth proceeded in a 0.02 Torr O$_2$ environment. Time-resolved X-ray CTR intensity measurements were made in the anti-Bragg scattering geometry in order to maximize sensitivity to surface roughness. PLD growth of $\langle 001 \rangle$ \sto\ in these conditions proceeds in a layer-by-layer growth mode, where many small two-dimensional islands nucleate, grow, and coalesce to form a complete overlayer. This type of growth gives rise to a strong discrete component of the surface roughness, which must be accounted for in order to fit the resulting oscillatory CTR intensity seen in the experimental data in figure \ref{fig:AB-oscillations}(a).

The parameterization of roughness contributions in equation \ref{eq:I-vs-sigma-general} enables a direct and simple method for analyzing scattering data from \textit{in-situ} studies of film growth. The binomial distribution is used here to describe the state of the surface, not the physics of the adsorption and diffusion processes. In place of $p$, the fractional coverage per pulse, we use $\theta$ to define the fractional coverage of the growing layer. For ideal layer-by-layer growth, we set $n=\sigma_d^2/\theta(1-\theta){c^\prime}^2=1$ always, and allow the coverage parameter $\theta$ to vary like a sawtooth waveform, where $0 \leq \theta \leq 1$. Since the scattered intensity depends on the quantity $\theta(1-\theta)$ and not $\theta$ itself, we can equivalently define $\theta$ as a triangular waveform
\begin{equation}
\label{eq:p-expansion}
\theta = \frac{1} {4} \left[ \frac{\sum_{j=0}^\infty \frac{\cos(2\pi j \langle k \rangle-\pi)} {(2j+1)^2}} {\sum_{j=0}^\infty\frac{1} {(2j+1)^2}} + 1 \right],
\end{equation}
\noindent where $0 \leq \theta \leq 0.5$ and $\langle k \rangle = \langle h \rangle/c^\prime$ is the normalized film thickness. In practice, neither layer completion nor half coverage occur at a specific moment across the entire sample, due to small position-dependent variations in the deposition rate. The series can be truncated at the first term to simulate this effect: $\theta=\cos(2\pi \langle k \rangle-\pi)/4.9352+1/4$.

As the film grows thicker and the continuous roughness increases due to the terraces become more and more irregularly shaped, we should expect that the influence of the discrete component will diminish. The decaying discrete roughness contribution is modeled using an effective coverage parameter $\theta'=\gamma \theta e^{-\langle k \rangle /\kappa}$, where $\gamma=0.8$ and $\kappa=18$. The presence of a characteristic thickness, $\kappa$, associated with the influence of discrete roughness provides a useful figure of merit for layer-by-layer growth.  Using scaling arguments described elsewhere,\cite{HN_Yang_1993, Barabasi_1995} we have modeled the continuous roughness as a power-law $\sigma_c=\alpha \langle k \rangle^{\beta}$, with $\alpha=0.617$ and $\beta=0.34$.  These models of the discrete, continuous, and total surface roughness as a function of film thickness is presented in figure \ref{fig:AB-oscillations}(b). By considering both the discrete and continuous components, we obtain good agreement between experimentally measured scattered intensity and the model as seen in figure \ref{fig:AB-oscillations}(a).

We have not observed a clear bimodal distribution in AFM data, even for post-deposition measurements of samples with many small two-dimensional islands or holes on the surface. In figure \ref{fig:STO-miscut}-(b), we show a hypothetical surface that exhibits a bimodal distribution, but we did not account for the effect of convoluting the AFM tip with the simulated morphology. Our AFM tips have a radius of approximately 10 nm, so it is not surprising that the AFM did not confirm the existence of the discrete roughness contribution. In this case, X-rays are more sensitive to such features.

\section{Conclusions}

We have used surface-sensitive, incoherent X-ray diffraction for time-resolved studies of $\langle 001 \rangle$ \sto\ surface morphology \textit{in-situ}. Quantitative analysis of specular Crystal Truncation Rod intensity measurements require a model that is capable of handling the oscillatory surface roughness that occurs during layer-by-layer growth, as well as the non-periodic roughness that continues to accumulate with increasing film thickness. In general, the RMS surface roughness alone is not sufficient to model Crystal Truncation Rod intensity. Crystals with small terrace size relative to the coherence length exhibit concomitant discrete and continuous surface roughness, which we have modeled as a convolution of continuous and discrete distributions. The total surface roughness can then be parameterized into continuous and discrete roughness contributions, which add in quadrature. We derived a simple, closed-form expression for the scattering intensity which, for specific limiting cases, is equivalent to other models existing in the literature. This model is capable of handling smooth transitions between roughness that is discrete in nature to roughness that is continuous in nature, and therefore relaxes some of the restrictions of existing models of reflectivity.

We have observed good agreement between this model and experimental observations, allowing quantitative investigations of $\langle 001 \rangle$ \sto\ annealing and thin film growth. We observed a surface roughening transition at temperatures that increased with decreasing O$_2$ pressure. The surface becomes smooth again as the temperature is further increased, and the rate of recovery was improved by increasing the O$_2$ pressure. The maximum surface roughness observed during the annealing process was 2.05\AA, and was primarily discrete in nature.

The continuous and discrete roughness parameters provide a useful method for analyzing \textit{in-situ} scattering measurements of thin film growth. Growth of oxide thin films via Pulsed Laser Deposition tend to proceed with a layer-by-layer mechanism, which gives rise to a strong discrete roughness component that is periodic with film thickness. As the film continues to grow, we have observed an increase in the continuous roughness as well, which may be due to increasing irregularity of the terraces or surface relaxation. As the continuous roughness increases, the influence of the discrete roughness decreases. The decay rate of the discrete roughness contribution may serve as a useful figure of merit for describing layer-by-layer growth.

\appendix
\section{}
\label{app:sum-overlayers}

We offer an alternate derivation for the effect of a binomial surface height distribution. The specularly scattered amplitude from a perfectly flat substrate can be written
\begin{equation}
\label{eq:A0}
A_0=\frac{F(Q_z)}{1-e^{iQ_z c_0}}.
\end{equation}
The scattering amplitude for an arbitrary surface, just before the $n^{\rm{th}}$ pulse, is
\begin{equation}
\label{eq:An-1}
A_{n-1}= A_0 + \sum_{k=1}^n \theta_k F(Q_z) e^{ -iQ_zc_0k },
\end{equation}
where $\theta_k$ is the fractional coverage of the $k^{th}$ overlayer. In the case of random deposition, the amount of material depositing onto layer $k-1$ is equal to $p(\theta_{k-1}-\theta_k)$, where $p$ is the fraction of the total surface covered by a single pulse. The scattering amplitude just after the $n^{\rm{th}}$ pulse is
\begin{eqnarray}
A_{n} & = & A_0 + \sum_{k=1}^n \left[ p \left( \theta_{k-1} - \theta_{k} \right) + \theta_k \right] F(Q_z) e^{-i Q_z c_0 k} \nonumber \\
& = & A_{n-1} - p \sum_{k=1}^n \theta_k F(Q_z) e^{ -iQ_z c_0 k } + p \sum_{k=1}^n {\theta_{k-1}} F(Q_z) e^{ -i Q_z c_0 k }
\label{eq:A-n+1}.
\end{eqnarray}
The second sum in equation \ref{eq:A-n+1} can be rewritten
$$
\sum_{k=1}^n {\theta_{k-1}} F(Q_z) e^{ -i Q_z c_0 k } = e^{ -i Q_z c_0 } \left[ F(Q_z) +\sum_{k=1}^n \theta_{k} F(Q_z) e^{ -i Q_z c_0 k } \right]
$$
such that, with some rearrangement, we can solve for $I_{n}$ in terms of $I_{n-1}$
\begin{eqnarray}
A_{n} & = & A_{n-1} - p \sum_{k=1}^n \theta_k F(Q_z) e^{ -iQ_z c_0 k } \nonumber \\
      &   & + p e^{ -i Q_z c_0 } \left[F(Q_z) \frac{1-e^{iQ_zc_0 }} {1-e^{iQ_zc_0}} + \sum_{k=1}^n \theta_{k} F(Q_z) e^{ -i Q_z c_0 k } \right] \nonumber \\
& = & A_{n-1} - p\left( 1-e^{-iQ_zc_0} \right)A_0 - p \left( 1-e^{-iQ_zc_0} \right) \sum_{k=1}^n \theta_{k} F(Q_z) e^{ -i Q_z c_0 k } \nonumber \\
& = & A_{n-1} \left( 1 - p + pe^{-iQ_zc_0} \right) \nonumber \\
I_n & = & I_{n-1}\left[ 1-4p(1-p)\sin^2\left( \frac{Q_zc_0}{2} \right) \right] \nonumber .
\end{eqnarray}
Starting from a perfect surface with no miscut, the intensity after $n$ pulses of random deposition, where the fraction $p_n$ of the surface covered during each pulse $p_n$ may vary, is therefore
\begin{equation}
I_n = I_0 \prod_n \left[ 1-4p_n(1-p_n)\sin^2\left( \frac{Q_zc_0}{2} \right) \right]
\label{eq:Intensity-n}.
\end{equation}

\begin{acknowledgments}
This work was funded by the Cornell Center for Materials Research (CCMR), which is supported by the National Science Foundation under award DMR-0079992, part of the NSF MRSEC Program. This work is based upon research conducted at the Cornell High Energy Synchrotron Source (CHESS) which is supported by the National Science Foundation and the National Institutes of Health/National Institute of General Medical Sciences under award DMR-0225180. Python and SciPy were used for data analysis, and Matplotlib was used to generate the figures herein. We wish to thank Arthur Woll for helpful discussions.
\end{acknowledgments}


\begin{thebibliography}{42}
\expandafter\ifx\csname natexlab\endcsname\relax\def\natexlab#1{#1}\fi
\expandafter\ifx\csname bibnamefont\endcsname\relax
  \def\bibnamefont#1{#1}\fi
\expandafter\ifx\csname bibfnamefont\endcsname\relax
  \def\bibfnamefont#1{#1}\fi
\expandafter\ifx\csname citenamefont\endcsname\relax
  \def\citenamefont#1{#1}\fi
\expandafter\ifx\csname url\endcsname\relax
  \def\url#1{\texttt{#1}}\fi
\expandafter\ifx\csname urlprefix\endcsname\relax\def\urlprefix{URL }\fi
\providecommand{\bibinfo}[2]{#2}
\providecommand{\eprint}[2][]{\url{#2}}

\bibitem[{\citenamefont{Stoop and Van
  Der~Merwe}(1973)}]{LCA_Stoop_1973_TSF_v17_p291}
\bibinfo{author}{\bibfnamefont{L.~C.~A.} \bibnamefont{Stoop}} \bibnamefont{and}
  \bibinfo{author}{\bibfnamefont{J.~H.} \bibnamefont{Van Der~Merwe}},
  \bibinfo{journal}{Thin Solid Films} \textbf{\bibinfo{volume}{17}},
  \bibinfo{pages}{291 } (\bibinfo{year}{1973}).

\bibitem[{\citenamefont{Frank}(1974)}]{FC_Frank_1974_JCG_v22_p233}
\bibinfo{author}{\bibfnamefont{F.}~\bibnamefont{Frank}}, \bibinfo{journal}{J.
  Cryst. Growth} \textbf{\bibinfo{volume}{22}}, \bibinfo{pages}{233 }
  (\bibinfo{year}{1974}).

\bibitem[{\citenamefont{Gilmer}(1980)}]{GH_Gilmer_1980_JCG_v49_p465}
\bibinfo{author}{\bibfnamefont{G.}~\bibnamefont{Gilmer}}, \bibinfo{journal}{J.
  Cryst. Growth} \textbf{\bibinfo{volume}{49}}, \bibinfo{pages}{465 }
  (\bibinfo{year}{1980}).

\bibitem[{\citenamefont{van Saarloos and
  Gilmer}(1986)}]{W_van_Saarloos_1986_PRB_v33_p4927}
\bibinfo{author}{\bibfnamefont{W.}~\bibnamefont{van Saarloos}}
  \bibnamefont{and} \bibinfo{author}{\bibfnamefont{G.~H.}~\bibnamefont{Gilmer}},
  \bibinfo{journal}{Phys. Rev. B} \textbf{\bibinfo{volume}{33}},
  \bibinfo{pages}{4927 } (\bibinfo{year}{1986}).

\bibitem[{\citenamefont{Ohtomo and
  Hwang}(2004)}]{A_Ohtomo_2004_Nature_v427_p423}
\bibinfo{author}{\bibfnamefont{A.}~\bibnamefont{Ohtomo}} \bibnamefont{and}
  \bibinfo{author}{\bibfnamefont{H.}~\bibnamefont{Hwang}},
  \bibinfo{journal}{Nature} \textbf{\bibinfo{volume}{427}},
  \bibinfo{pages}{423} (\bibinfo{year}{2004}).

\bibitem[{\citenamefont{Terashima et~al.}(1990)\citenamefont{Terashima, Bando,
  Iijima, Yamamoto, Hirata, Hayashi, Kamigaki, and
  Terauchi}}]{T_Terashima_1990_PRL_v65_p2684}
\bibinfo{author}{\bibfnamefont{T.}~\bibnamefont{Terashima}},
  \bibinfo{author}{\bibfnamefont{Y.}~\bibnamefont{Bando}},
  \bibinfo{author}{\bibfnamefont{K.}~\bibnamefont{Iijima}},
  \bibinfo{author}{\bibfnamefont{K.}~\bibnamefont{Yamamoto}},
  \bibinfo{author}{\bibfnamefont{K.}~\bibnamefont{Hirata}},
  \bibinfo{author}{\bibfnamefont{K.}~\bibnamefont{Hayashi}},
  \bibinfo{author}{\bibfnamefont{K.}~\bibnamefont{Kamigaki}}, \bibnamefont{and}
  \bibinfo{author}{\bibfnamefont{H.}~\bibnamefont{Terauchi}},
  \bibinfo{journal}{Phys. Rev. Lett.} \textbf{\bibinfo{volume}{65}},
  \bibinfo{pages}{2684 } (\bibinfo{year}{1990}).

\bibitem[{\citenamefont{Karl and Stritzker}(1992)}]{H_Karl_1992_PRL_v69_p2939}
\bibinfo{author}{\bibfnamefont{H.}~\bibnamefont{Karl}} \bibnamefont{and}
  \bibinfo{author}{\bibfnamefont{B.}~\bibnamefont{Stritzker}},
  \bibinfo{journal}{Phys. Rev. Lett.} \textbf{\bibinfo{volume}{69}},
  \bibinfo{pages}{2939 } (\bibinfo{year}{1992}).

\bibitem[{\citenamefont{Koster et~al.}(1999)\citenamefont{Koster, Rijnders,
  Blank, and Rogalla}}]{G_Koster_1999_APL_v74_p3729}
\bibinfo{author}{\bibfnamefont{G.}~\bibnamefont{Koster}},
  \bibinfo{author}{\bibfnamefont{G.}~\bibnamefont{Rijnders}},
  \bibinfo{author}{\bibfnamefont{D.}~\bibnamefont{Blank}}, \bibnamefont{and}
  \bibinfo{author}{\bibfnamefont{H.}~\bibnamefont{Rogalla}},
  \bibinfo{journal}{Appl. Phys. Lett.} \textbf{\bibinfo{volume}{74}},
  \bibinfo{pages}{3729} (\bibinfo{year}{1999}).

\bibitem[{\citenamefont{Ohtomo et~al.}(2002)\citenamefont{Ohtomo, Muller,
  Grazul, and Hwang}}]{A_Ohtomo_2002_Nature_v419_p378}
\bibinfo{author}{\bibfnamefont{A.}~\bibnamefont{Ohtomo}},
  \bibinfo{author}{\bibfnamefont{D.}~\bibnamefont{Muller}},
  \bibinfo{author}{\bibfnamefont{J.}~\bibnamefont{Grazul}}, \bibnamefont{and}
  \bibinfo{author}{\bibfnamefont{H.}~\bibnamefont{Hwang}},
  \bibinfo{journal}{Nature} \textbf{\bibinfo{volume}{419}},
  \bibinfo{pages}{378} (\bibinfo{year}{2002}).

\bibitem[{\citenamefont{Fleet et~al.}(2005)\citenamefont{Fleet, Dale, Suzuki,
  and Brock}}]{A_Fleet_2005_PRL_v94_p036102}
\bibinfo{author}{\bibfnamefont{A.}~\bibnamefont{Fleet}},
  \bibinfo{author}{\bibfnamefont{D.}~\bibnamefont{Dale}},
  \bibinfo{author}{\bibfnamefont{Y.}~\bibnamefont{Suzuki}}, \bibnamefont{and}
  \bibinfo{author}{\bibfnamefont{J.~D.}~\bibnamefont{Brock}},
  \bibinfo{journal}{Phys. Rev. Lett.} \textbf{\bibinfo{volume}{94}},
  \bibinfo{pages}{036102} (\bibinfo{year}{2005}).

\bibitem[{\citenamefont{Fleet et~al.}(2006)\citenamefont{Fleet, Dale, Woll,
  Suzuki, and Brock}}]{A_Fleet_2006_PRL_v96_p055508}
\bibinfo{author}{\bibfnamefont{A.}~\bibnamefont{Fleet}},
  \bibinfo{author}{\bibfnamefont{D.}~\bibnamefont{Dale}},
  \bibinfo{author}{\bibfnamefont{A.~R.} \bibnamefont{Woll}},
  \bibinfo{author}{\bibfnamefont{Y.}~\bibnamefont{Suzuki}}, \bibnamefont{and}
  \bibinfo{author}{\bibfnamefont{J.~D.} \bibnamefont{Brock}},
  \bibinfo{journal}{Phys. Rev. Lett.} \textbf{\bibinfo{volume}{96}},
  \bibinfo{eid}{055508} (\bibinfo{year}{2006}).

\bibitem[{\citenamefont{Kisker et~al.}(1996)\citenamefont{Kisker, Stephenson,
  Tersoff, Fuoss, and Brennan}}]{DW_Kisker_1996_JCG_v163_p54}
\bibinfo{author}{\bibfnamefont{D.~W.} \bibnamefont{Kisker}},
  \bibinfo{author}{\bibfnamefont{G.~B.} \bibnamefont{Stephenson}},
  \bibinfo{author}{\bibfnamefont{J.}~\bibnamefont{Tersoff}},
  \bibinfo{author}{\bibfnamefont{P.~H.} \bibnamefont{Fuoss}}, \bibnamefont{and}
  \bibinfo{author}{\bibfnamefont{S.}~\bibnamefont{Brennan}},
  \bibinfo{journal}{J. Cryst. Growth} \textbf{\bibinfo{volume}{163}},
  \bibinfo{pages}{54} (\bibinfo{year}{1996}).

\bibitem[{\citenamefont{Headrick et~al.}(1996)\citenamefont{Headrick, Kycia,
  Park, Woll, and Brock}}]{RL_Headrick_1996_PRB_v54_p14686}
\bibinfo{author}{\bibfnamefont{R.~L.}~\bibnamefont{Headrick}},
  \bibinfo{author}{\bibfnamefont{S.}~\bibnamefont{Kycia}},
  \bibinfo{author}{\bibfnamefont{Y.~K.}~\bibnamefont{Park}},
  \bibinfo{author}{\bibfnamefont{A.~R.}~\bibnamefont{Woll}}, \bibnamefont{and}
  \bibinfo{author}{\bibfnamefont{J.~D.}~\bibnamefont{Brock}},
  \bibinfo{journal}{Phys. Rev. B} \textbf{\bibinfo{volume}{54}},
  \bibinfo{pages}{14686 } (\bibinfo{year}{1996}).

\bibitem[{\citenamefont{Ramana~Murty et~al.}(1999)\citenamefont{Ramana~Murty,
  Curcic, Judy, Cooper, Woll, Brock, Kycia, and
  Headrick}}]{MV_RamanaMurty_1999_PRB_v60_p16956}
\bibinfo{author}{\bibfnamefont{M.~V.}~\bibnamefont{Ramana~Murty}},
  \bibinfo{author}{\bibfnamefont{T.}~\bibnamefont{Curcic}},
  \bibinfo{author}{\bibfnamefont{A.}~\bibnamefont{Judy}},
  \bibinfo{author}{\bibfnamefont{B.~H.}~\bibnamefont{Cooper}},
  \bibinfo{author}{\bibfnamefont{A.~R.}~\bibnamefont{Woll}},
  \bibinfo{author}{\bibfnamefont{J.~D.}~\bibnamefont{Brock}},
  \bibinfo{author}{\bibfnamefont{S.}~\bibnamefont{Kycia}}, \bibnamefont{and}
  \bibinfo{author}{\bibfnamefont{R.~L.}~\bibnamefont{Headrick}},
  \bibinfo{journal}{Phys. Rev. B} \textbf{\bibinfo{volume}{60}},
  \bibinfo{pages}{16956 } (\bibinfo{year}{1999}).

\bibitem[{\citenamefont{Eres et~al.}(2002)\citenamefont{Eres, Tischler, Yoon,
  Larson, Rouleau, Lowndes, and Zschack}}]{G_Eres_2002_APL_v80_p3379}
\bibinfo{author}{\bibfnamefont{G.}~\bibnamefont{Eres}},
  \bibinfo{author}{\bibfnamefont{J.~Z.} \bibnamefont{Tischler}},
  \bibinfo{author}{\bibfnamefont{M.}~\bibnamefont{Yoon}},
  \bibinfo{author}{\bibfnamefont{B.~C.} \bibnamefont{Larson}},
  \bibinfo{author}{\bibfnamefont{C.~M.} \bibnamefont{Rouleau}},
  \bibinfo{author}{\bibfnamefont{D.~H.} \bibnamefont{Lowndes}},
  \bibnamefont{and} \bibinfo{author}{\bibfnamefont{P.}~\bibnamefont{Zschack}},
  \bibinfo{journal}{Appl. Phys. Lett.} \textbf{\bibinfo{volume}{80}},
  \bibinfo{pages}{3379} (\bibinfo{year}{2002}).

\bibitem[{\citenamefont{Vlieg et~al.}(1988)\citenamefont{Vlieg, Denier van~der
  Gon, van~der Veen, Macdonald, and Norris}}]{E_Vlieg_1988_PRL_v61_p2241}
\bibinfo{author}{\bibfnamefont{E.}~\bibnamefont{Vlieg}},
  \bibinfo{author}{\bibfnamefont{A.~W.}~\bibnamefont{Denier van~der Gon}},
  \bibinfo{author}{\bibfnamefont{J.~F.}~\bibnamefont{van~der Veen}},
  \bibinfo{author}{\bibfnamefont{J.~E.}~\bibnamefont{Macdonald}},
  \bibnamefont{and} \bibinfo{author}{\bibfnamefont{C.}~\bibnamefont{Norris}},
  \bibinfo{journal}{Phys. Rev. Lett.} \textbf{\bibinfo{volume}{61}},
  \bibinfo{pages}{2241 } (\bibinfo{year}{1988}).

\bibitem[{\citenamefont{Fuoss et~al.}(1992)\citenamefont{Fuoss, Kisker,
  Lamelas, Stephenson, Imperatori, and Brennan}}]{PH_Fuoss_1992_PRL_v69_p2791}
\bibinfo{author}{\bibfnamefont{P.~H.} \bibnamefont{Fuoss}},
  \bibinfo{author}{\bibfnamefont{D.~W.} \bibnamefont{Kisker}},
  \bibinfo{author}{\bibfnamefont{F.~J.} \bibnamefont{Lamelas}},
  \bibinfo{author}{\bibfnamefont{G.~B.} \bibnamefont{Stephenson}},
  \bibinfo{author}{\bibfnamefont{P.}~\bibnamefont{Imperatori}},
  \bibnamefont{and} \bibinfo{author}{\bibfnamefont{S.}~\bibnamefont{Brennan}},
  \bibinfo{journal}{Phys. Rev. Lett.} \textbf{\bibinfo{volume}{69}},
  \bibinfo{pages}{2791} (\bibinfo{year}{1992}).

\bibitem[{\citenamefont{Woll et~al.}(1999)\citenamefont{Woll, Headrick, Kycia,
  and Brock}}]{AR_Woll_1999_PRL_v83_p4349}
\bibinfo{author}{\bibfnamefont{A.~R.}~\bibnamefont{Woll}},
  \bibinfo{author}{\bibfnamefont{R.~L.}~\bibnamefont{Headrick}},
  \bibinfo{author}{\bibfnamefont{S.}~\bibnamefont{Kycia}}, \bibnamefont{and}
  \bibinfo{author}{\bibfnamefont{J.~D.}~\bibnamefont{Brock}},
  \bibinfo{journal}{Phys. Rev. Lett.} \textbf{\bibinfo{volume}{83}},
  \bibinfo{pages}{4349 } (\bibinfo{year}{1999}).

\bibitem[{\citenamefont{Willmott et~al.}(2006)\citenamefont{Willmott, Herger,
  Schleputz, Martoccia, and Patterson}}]{PR_Willmott_2006_PRL_v96_p176102}
\bibinfo{author}{\bibfnamefont{P.~R.} \bibnamefont{Willmott}},
  \bibinfo{author}{\bibfnamefont{R.}~\bibnamefont{Herger}},
  \bibinfo{author}{\bibfnamefont{C.~M.} \bibnamefont{Schleputz}},
  \bibinfo{author}{\bibfnamefont{D.}~\bibnamefont{Martoccia}},
  \bibnamefont{and} \bibinfo{author}{\bibfnamefont{B.~D.}
  \bibnamefont{Patterson}}, \bibinfo{journal}{Phys. Rev. Lett.}
  \textbf{\bibinfo{volume}{96}}, \bibinfo{pages}{176102}
  (\bibinfo{year}{2006}).

\bibitem[{\citenamefont{Tischler et~al.}(2006)\citenamefont{Tischler, Eres,
  Larson, Rouleau, Zschack, and Lowndes}}]{JZ_Tischler_2006_PRL_v96_p226104}
\bibinfo{author}{\bibfnamefont{J.~Z.}~\bibnamefont{Tischler}},
  \bibinfo{author}{\bibfnamefont{G.}~\bibnamefont{Eres}},
  \bibinfo{author}{\bibfnamefont{B.~C.}~\bibnamefont{Larson}},
  \bibinfo{author}{\bibfnamefont{C.~M.} \bibnamefont{Rouleau}},
  \bibinfo{author}{\bibfnamefont{P.}~\bibnamefont{Zshack}}, \bibnamefont{and}
  \bibinfo{author}{\bibfnamefont{D.~H.} \bibnamefont{Lowndes}},
  \bibinfo{journal}{Phys. Rev. Lett.} \textbf{\bibinfo{volume}{96}},
  \bibinfo{pages}{226104} (\bibinfo{year}{2006}).

\bibitem[{\citenamefont{Iijima et~al.}(1992)\citenamefont{Iijima, Terashima,
  Bando, Kamigaki, and Terauchi}}]{K_Iijima_1992_JAP_v72_p2840}
\bibinfo{author}{\bibfnamefont{K.}~\bibnamefont{Iijima}},
  \bibinfo{author}{\bibfnamefont{T.}~\bibnamefont{Terashima}},
  \bibinfo{author}{\bibfnamefont{Y.}~\bibnamefont{Bando}},
  \bibinfo{author}{\bibfnamefont{K.}~\bibnamefont{Kamigaki}}, \bibnamefont{and}
  \bibinfo{author}{\bibfnamefont{H.}~\bibnamefont{Terauchi}},
  \bibinfo{journal}{J. Appl. Phys.} \textbf{\bibinfo{volume}{72}},
  \bibinfo{pages}{2840 } (\bibinfo{year}{1992}).

\bibitem[{\citenamefont{Tsurumi et~al.}(1994)\citenamefont{Tsurumi, Suzuki,
  Yamane, and Daimon}}]{T_Tsurumi_1994_JJAP_v33_p5192}
\bibinfo{author}{\bibfnamefont{T.}~\bibnamefont{Tsurumi}},
  \bibinfo{author}{\bibfnamefont{T.}~\bibnamefont{Suzuki}},
  \bibinfo{author}{\bibfnamefont{M.}~\bibnamefont{Yamane}}, \bibnamefont{and}
  \bibinfo{author}{\bibfnamefont{M.}~\bibnamefont{Daimon}},
  \bibinfo{journal}{Jpn. J. Appl. Phys.} \textbf{\bibinfo{volume}{33}},
  \bibinfo{pages}{5192 } (\bibinfo{year}{1994}).

\bibitem[{\citenamefont{Tabata et~al.}(1994)\citenamefont{Tabata, Tanaka, and
  Kawai}}]{H_Tabata_1994_APL_v65_p1970}
\bibinfo{author}{\bibfnamefont{H.}~\bibnamefont{Tabata}},
  \bibinfo{author}{\bibfnamefont{H.}~\bibnamefont{Tanaka}}, \bibnamefont{and}
  \bibinfo{author}{\bibfnamefont{T.}~\bibnamefont{Kawai}},
  \bibinfo{journal}{Appl. Phys. Lett.} \textbf{\bibinfo{volume}{65}},
  \bibinfo{pages}{1970 } (\bibinfo{year}{1994}).

\bibitem[{\citenamefont{Takahashi et~al.}(2002)\citenamefont{Takahashi,
  Matsumoto, Koinuma, Lippmaa, and Kawasaki}}]{R_Takahashi_2002_ASS_v197_p532}
\bibinfo{author}{\bibfnamefont{R.}~\bibnamefont{Takahashi}},
  \bibinfo{author}{\bibfnamefont{Y.}~\bibnamefont{Matsumoto}},
  \bibinfo{author}{\bibfnamefont{H.}~\bibnamefont{Koinuma}},
  \bibinfo{author}{\bibfnamefont{M.}~\bibnamefont{Lippmaa}}, \bibnamefont{and}
  \bibinfo{author}{\bibfnamefont{M.}~\bibnamefont{Kawasaki}},
  \bibinfo{journal}{Appl. Surf. Sci.} \textbf{\bibinfo{volume}{197}},
  \bibinfo{pages}{532} (\bibinfo{year}{2002}).

\bibitem[{\citenamefont{Koster et~al.}(2001)\citenamefont{Koster, Verbist,
  Rijnders, Rogalla, van Tendeloo, and Blank}}]{G_Koster_2001_PC_v353_p167}
\bibinfo{author}{\bibfnamefont{G.}~\bibnamefont{Koster}},
  \bibinfo{author}{\bibfnamefont{K.}~\bibnamefont{Verbist}},
  \bibinfo{author}{\bibfnamefont{G.}~\bibnamefont{Rijnders}},
  \bibinfo{author}{\bibfnamefont{H.}~\bibnamefont{Rogalla}},
  \bibinfo{author}{\bibfnamefont{G.}~\bibnamefont{van Tendeloo}},
  \bibnamefont{and} \bibinfo{author}{\bibfnamefont{D.}~\bibnamefont{Blank}},
  \bibinfo{journal}{Physica C} \textbf{\bibinfo{volume}{353}},
  \bibinfo{pages}{167} (\bibinfo{year}{2001}).

\bibitem[{\citenamefont{Tsurumi et~al.}(2004)\citenamefont{Tsurumi, Harigai,
  Tanaka, Nam, Kakemoto, Wada, and Saito}}]{T_Tsurumi_2004_APL_v85_p5016}
\bibinfo{author}{\bibfnamefont{T.}~\bibnamefont{Tsurumi}},
  \bibinfo{author}{\bibfnamefont{T.}~\bibnamefont{Harigai}},
  \bibinfo{author}{\bibfnamefont{D.}~\bibnamefont{Tanaka}},
  \bibinfo{author}{\bibfnamefont{S.-M.} \bibnamefont{Nam}},
  \bibinfo{author}{\bibfnamefont{H.}~\bibnamefont{Kakemoto}},
  \bibinfo{author}{\bibfnamefont{S.}~\bibnamefont{Wada}}, \bibnamefont{and}
  \bibinfo{author}{\bibfnamefont{K.}~\bibnamefont{Saito}},
  \bibinfo{journal}{Appl. Phys. Lett.} \textbf{\bibinfo{volume}{85}},
  \bibinfo{pages}{5016 } (\bibinfo{year}{2004}).

\bibitem[{\citenamefont{Lee et~al.}(2005)\citenamefont{Lee, Christen, Chisholm,
  Rouleau, and Lowndes}}]{HN_Lee_2005_Nature_v433_p395}
\bibinfo{author}{\bibfnamefont{H.~N.} \bibnamefont{Lee}},
  \bibinfo{author}{\bibfnamefont{H.}~\bibnamefont{Christen}},
  \bibinfo{author}{\bibfnamefont{M.}~\bibnamefont{Chisholm}},
  \bibinfo{author}{\bibfnamefont{C.}~\bibnamefont{Rouleau}}, \bibnamefont{and}
  \bibinfo{author}{\bibfnamefont{D.}~\bibnamefont{Lowndes}},
  \bibinfo{journal}{Nature} \textbf{\bibinfo{volume}{433}}, \bibinfo{pages}{395
  } (\bibinfo{year}{2005}).

\bibitem[{\citenamefont{Vidal and Vincent}(1984)}]{B_Vidal_1984_AO_v23_p1794}
\bibinfo{author}{\bibfnamefont{B.}~\bibnamefont{Vidal}} \bibnamefont{and}
  \bibinfo{author}{\bibfnamefont{P.}~\bibnamefont{Vincent}},
  \bibinfo{journal}{Applied Optics} \textbf{\bibinfo{volume}{23}},
  \bibinfo{pages}{1794} (\bibinfo{year}{1984}).

\bibitem[{\citenamefont{Lent and Cohen}(1984)}]{CS_Lent_1984_SS_v139_p121}
\bibinfo{author}{\bibfnamefont{C.}~\bibnamefont{Lent}} \bibnamefont{and}
  \bibinfo{author}{\bibfnamefont{P.}~\bibnamefont{Cohen}},
  \bibinfo{journal}{Surf. Sci.} \textbf{\bibinfo{volume}{139}},
  \bibinfo{pages}{121 } (\bibinfo{year}{1984}).

\bibitem[{\citenamefont{Yang et~al.}(1993)\citenamefont{Yang, Wang, and
  Lu}}]{HN_Yang_1993}
\bibinfo{author}{\bibfnamefont{H.-N.} \bibnamefont{Yang}},
  \bibinfo{author}{\bibfnamefont{G.-C.} \bibnamefont{Wang}}, \bibnamefont{and}
  \bibinfo{author}{\bibfnamefont{T.-M.} \bibnamefont{Lu}},
  \emph{\bibinfo{title}{Diffraction from Rough Surfaces and Dynamic Growth
  Fronts}} (\bibinfo{publisher}{World Scientific Publishing Co.},
  \bibinfo{year}{1993}), \bibinfo{note}{and references therein}.

\bibitem[{\citenamefont{Gibbs et~al.}(1988)\citenamefont{Gibbs, Ocko, Zehner,
  and Mochrie}}]{D_Gibbs_1988_PRB_v38_p7303}
\bibinfo{author}{\bibfnamefont{D.}~\bibnamefont{Gibbs}},
  \bibinfo{author}{\bibfnamefont{B.~M.}~\bibnamefont{Ocko}},
  \bibinfo{author}{\bibfnamefont{D.~M.}~\bibnamefont{Zehner}}, \bibnamefont{and}
  \bibinfo{author}{\bibfnamefont{S.~G.~J.}~\bibnamefont{Mochrie}},
  \bibinfo{journal}{Phys. Rev. B} \textbf{\bibinfo{volume}{38}},
  \bibinfo{pages}{7303} (\bibinfo{year}{1988}).

\bibitem[{\citenamefont{Specht and Walker}(1993)}]{ED_Specht_1993_JAC_v26_p166}
\bibinfo{author}{\bibfnamefont{E.~D.} \bibnamefont{Specht}} \bibnamefont{and}
  \bibinfo{author}{\bibfnamefont{F.~J.} \bibnamefont{Walker}},
  \bibinfo{journal}{J. Appl. Crystallogr.} \textbf{\bibinfo{volume}{26}},
  \bibinfo{pages}{166} (\bibinfo{year}{1993}).

\bibitem[{\citenamefont{Warren}(1990)}]{Warren_1990}
\bibinfo{author}{\bibfnamefont{B.~E.} \bibnamefont{Warren}},
  \emph{\bibinfo{title}{X-ray Diffraction}} (\bibinfo{publisher}{Dover
  Publications}, \bibinfo{year}{1990}).

\bibitem[{\citenamefont{Sinha et~al.}(1988)\citenamefont{Sinha, Sirota, Garoff,
  and Stanley}}]{SK_Sinha_1988_PRB_v38_p2297}
\bibinfo{author}{\bibfnamefont{S.~K.}~\bibnamefont{Sinha}},
  \bibinfo{author}{\bibfnamefont{E.~B.}~\bibnamefont{Sirota}},
  \bibinfo{author}{\bibfnamefont{S.}~\bibnamefont{Garoff}}, \bibnamefont{and}
  \bibinfo{author}{\bibfnamefont{H.~B.}~\bibnamefont{Stanley}},
  \bibinfo{journal}{Phys. Rev. B} \textbf{\bibinfo{volume}{38}},
  \bibinfo{pages}{2297 } (\bibinfo{year}{1988}).

\bibitem[{\citenamefont{Held and Brock}(1995)}]{G_Held_1995_PRB_v51_p7262}
\bibinfo{author}{\bibfnamefont{G.~A.}~\bibnamefont{Held}} \bibnamefont{and}
  \bibinfo{author}{\bibfnamefont{J.~D.}~\bibnamefont{Brock}},
  \bibinfo{journal}{Phys. Rev. B} \textbf{\bibinfo{volume}{51}},
  \bibinfo{pages}{7262 } (\bibinfo{year}{1995}).

\bibitem[{\citenamefont{Pimbley and Lu}(1984)}]{JM_Pimbley_1984_JVSTA_v2_p457}
\bibinfo{author}{\bibfnamefont{J.}~\bibnamefont{Pimbley}} \bibnamefont{and}
  \bibinfo{author}{\bibfnamefont{T.-M.} \bibnamefont{Lu}}, \bibinfo{journal}{J.
  Vac. Sci. Technol. A} \textbf{\bibinfo{volume}{2}}, \bibinfo{pages}{457 }
  (\bibinfo{year}{1984}).

\bibitem[{\citenamefont{Andrews and
  Cowley}(1985)}]{SR_Andrews_1985_JPC_v18_p6427}
\bibinfo{author}{\bibfnamefont{S.}~\bibnamefont{Andrews}} \bibnamefont{and}
  \bibinfo{author}{\bibfnamefont{R.}~\bibnamefont{Cowley}},
  \bibinfo{journal}{J. Phys. C} \textbf{\bibinfo{volume}{18}},
  \bibinfo{pages}{6427 } (\bibinfo{year}{1985}).

\bibitem[{\citenamefont{Sinha et~al.}(1996)\citenamefont{Sinha, Feng,
  Melendres, Lee, Russell, Satija, Sirota, and
  Sanyal}}]{SK_Sinha_1996_PA_v231_p99}
\bibinfo{author}{\bibfnamefont{S.}~\bibnamefont{Sinha}},
  \bibinfo{author}{\bibfnamefont{Y.}~\bibnamefont{Feng}},
  \bibinfo{author}{\bibfnamefont{C.}~\bibnamefont{Melendres}},
  \bibinfo{author}{\bibfnamefont{D.}~\bibnamefont{Lee}},
  \bibinfo{author}{\bibfnamefont{T.}~\bibnamefont{Russell}},
  \bibinfo{author}{\bibfnamefont{S.}~\bibnamefont{Satija}},
  \bibinfo{author}{\bibfnamefont{E.}~\bibnamefont{Sirota}}, \bibnamefont{and}
  \bibinfo{author}{\bibfnamefont{M.}~\bibnamefont{Sanyal}},
  \bibinfo{journal}{Physica A} \textbf{\bibinfo{volume}{231}},
  \bibinfo{pages}{99 } (\bibinfo{year}{1996}).

\bibitem[{\citenamefont{Robinson}(1986)}]{IK_Robinson_1986_PRB_v33_p3830}
\bibinfo{author}{\bibfnamefont{I.~K.} \bibnamefont{Robinson}},
  \bibinfo{journal}{Phys. Rev. B} \textbf{\bibinfo{volume}{33}},
  \bibinfo{pages}{3830} (\bibinfo{year}{1986}).

\bibitem[{\citenamefont{Kawasaki et~al.}(1994)\citenamefont{Kawasaki,
  Takahashi, Maeda, Tsuchiya, Shinohara, Ishiyama, Yonezawa, Yoshimoto, and
  Koinuma}}]{M_Kawasaki_1994_Science_v266_p1540}
\bibinfo{author}{\bibfnamefont{M.}~\bibnamefont{Kawasaki}},
  \bibinfo{author}{\bibfnamefont{K.}~\bibnamefont{Takahashi}},
  \bibinfo{author}{\bibfnamefont{T.}~\bibnamefont{Maeda}},
  \bibinfo{author}{\bibfnamefont{R.}~\bibnamefont{Tsuchiya}},
  \bibinfo{author}{\bibfnamefont{M.}~\bibnamefont{Shinohara}},
  \bibinfo{author}{\bibfnamefont{O.}~\bibnamefont{Ishiyama}},
  \bibinfo{author}{\bibfnamefont{T.}~\bibnamefont{Yonezawa}},
  \bibinfo{author}{\bibfnamefont{M.}~\bibnamefont{Yoshimoto}},
  \bibnamefont{and} \bibinfo{author}{\bibfnamefont{H.}~\bibnamefont{Koinuma}},
  \bibinfo{journal}{Science} \textbf{\bibinfo{volume}{266}},
  \bibinfo{pages}{1540 } (\bibinfo{year}{1994}).

\bibitem[{\citenamefont{Lippmaa et~al.}(1998)\citenamefont{Lippmaa, Kawasaki,
  Ohtomo, Sato, Iwatsuki, and Koinuma}}]{M_Lippmaa_1998_ASS_v130_p582}
\bibinfo{author}{\bibfnamefont{M.}~\bibnamefont{Lippmaa}},
  \bibinfo{author}{\bibfnamefont{M.}~\bibnamefont{Kawasaki}},
  \bibinfo{author}{\bibfnamefont{A.}~\bibnamefont{Ohtomo}},
  \bibinfo{author}{\bibfnamefont{T.}~\bibnamefont{Sato}},
  \bibinfo{author}{\bibfnamefont{M.}~\bibnamefont{Iwatsuki}}, \bibnamefont{and}
  \bibinfo{author}{\bibfnamefont{H.}~\bibnamefont{Koinuma}},
  \bibinfo{journal}{Appl. Surf. Sci.} \textbf{\bibinfo{volume}{132}},
  \bibinfo{pages}{582} (\bibinfo{year}{1998}).

\bibitem[{\citenamefont{Barabasi and Stanley}(1995)}]{Barabasi_1995}
\bibinfo{author}{\bibfnamefont{A.-L.} \bibnamefont{Barabasi}} \bibnamefont{and}
  \bibinfo{author}{\bibfnamefont{H.~E.} \bibnamefont{Stanley}},
  \emph{\bibinfo{title}{Fractal Concepts in Surface Growth}}
  (\bibinfo{publisher}{Cambridge University Press}, \bibinfo{year}{1995}).

\end{thebibliography}

\begin{figure}[tb!]
    \centering
    \includegraphics[clip=true]{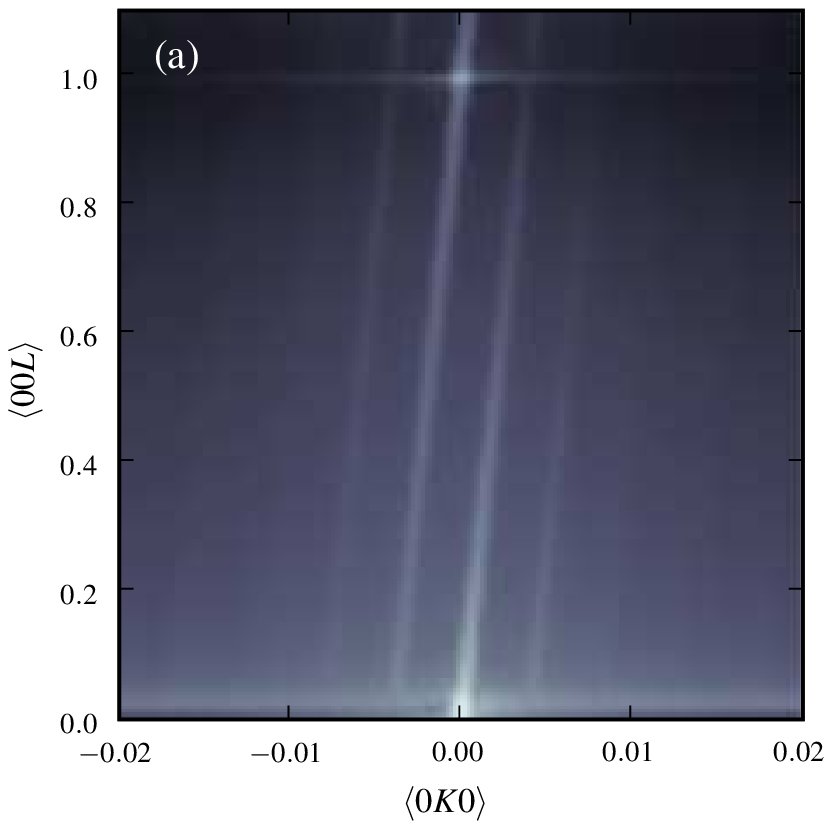}
    \includegraphics[clip=true]{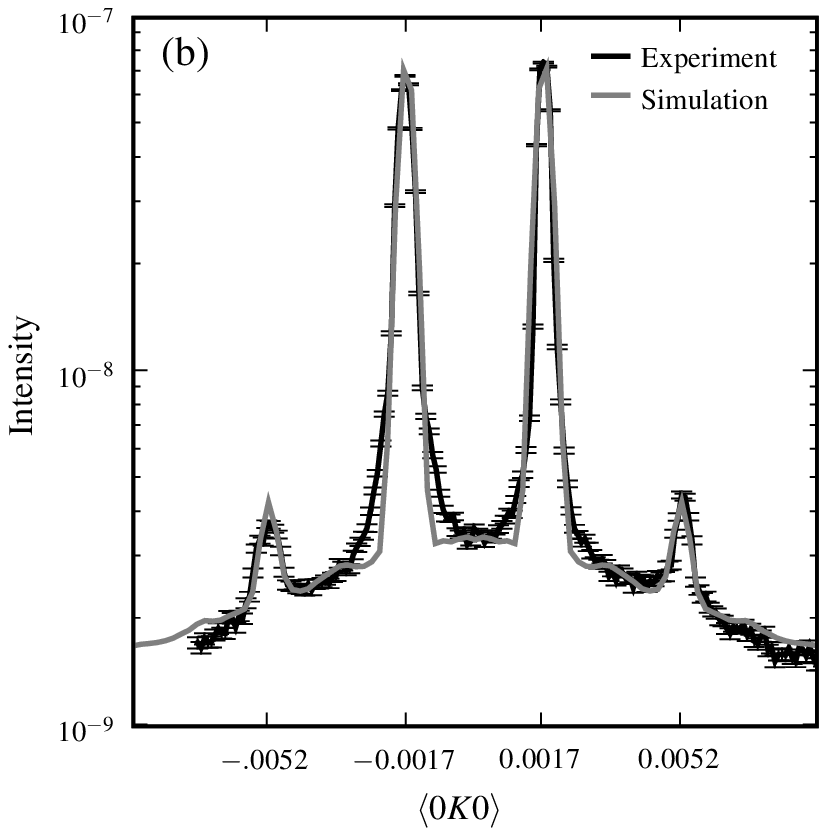}
    \caption{Darren Dale, Physical Review B}
    \label{fig:STO-rsmap}
\end{figure}

\clearpage
Figure \ref{fig:STO-rsmap}: (a) Simulated reciprocal space map ($H=0$) of intensity for a TiO$_2$ terminated SrTiO$_3$ substrate with 0.198$^\circ$ miscut in the $\langle 0K0 \rangle$ direction. The transverse coherence length is assumed to be $1\mu$m. (b) Experimental $K$-scan data from a $\langle 001 \rangle$ SrTiO$_3$ substrate, $H=0, L=0.5$. The scattering power is normalized to the total incident power. The grey line is simulated with the kinematic scattering theory.

\begin{figure}[tb!]
    \centering
    \includegraphics[clip=true]{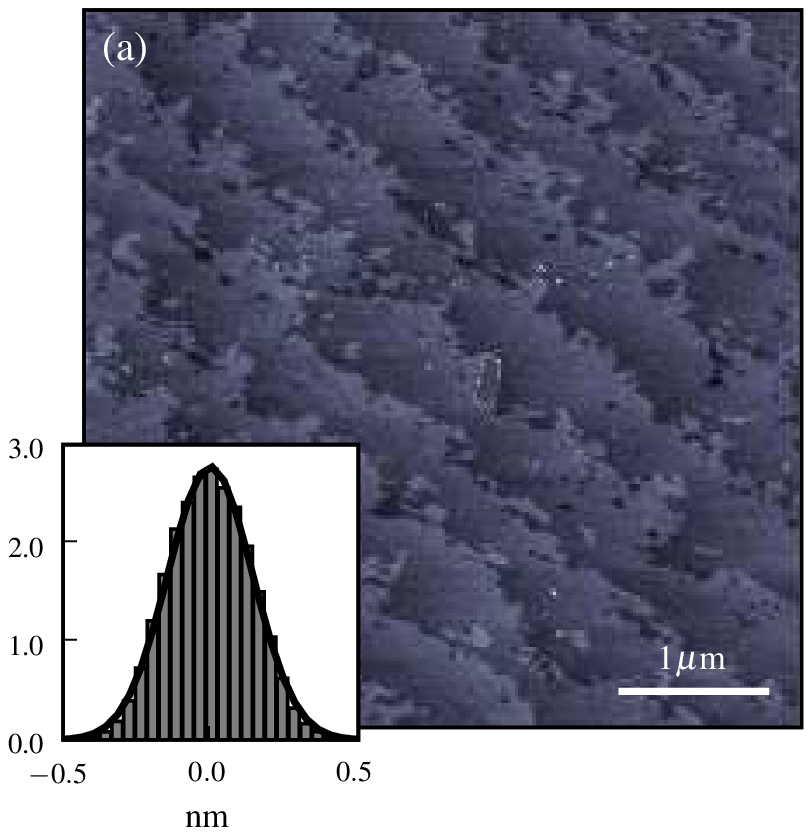}
    \includegraphics[clip=true]{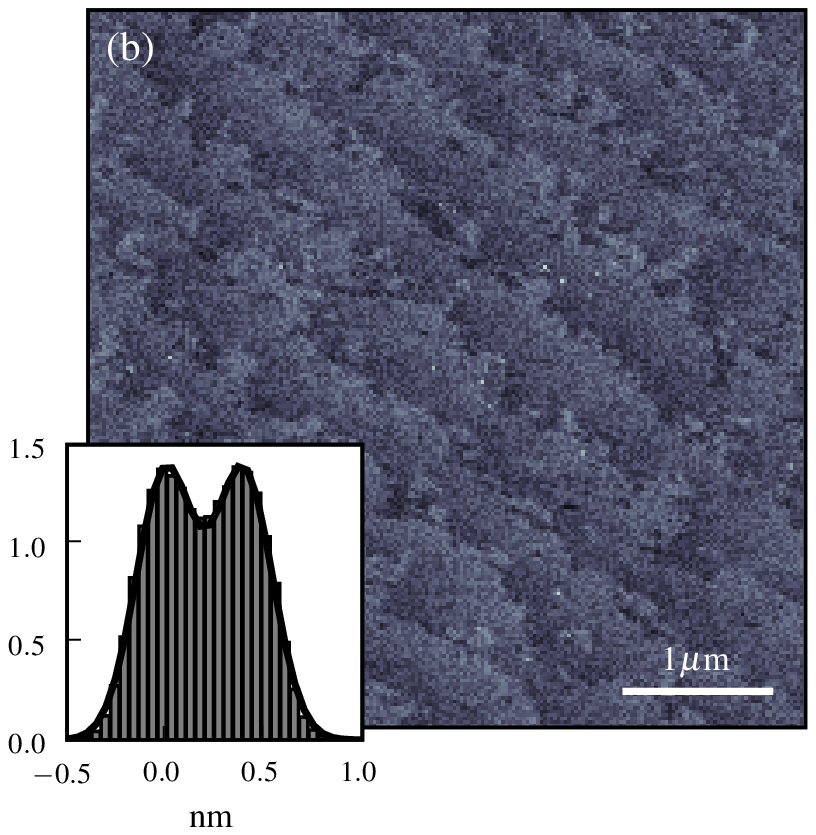}
    \caption{Darren Dale, Physical Review B}
    \label{fig:STO-miscut}
\end{figure}

Figure \ref{fig:STO-miscut}: (a) AFM image of a SrTiO$_3$ substrate. Inset: Histogram of the surface height distribution, with a Gaussian fit. The RMS roughness is 1.44\ \AA. (b) Using image (a), simulation of random deposition of one half monolayer.

\begin{figure}[tb!]
    \centering
    \includegraphics[clip=true] {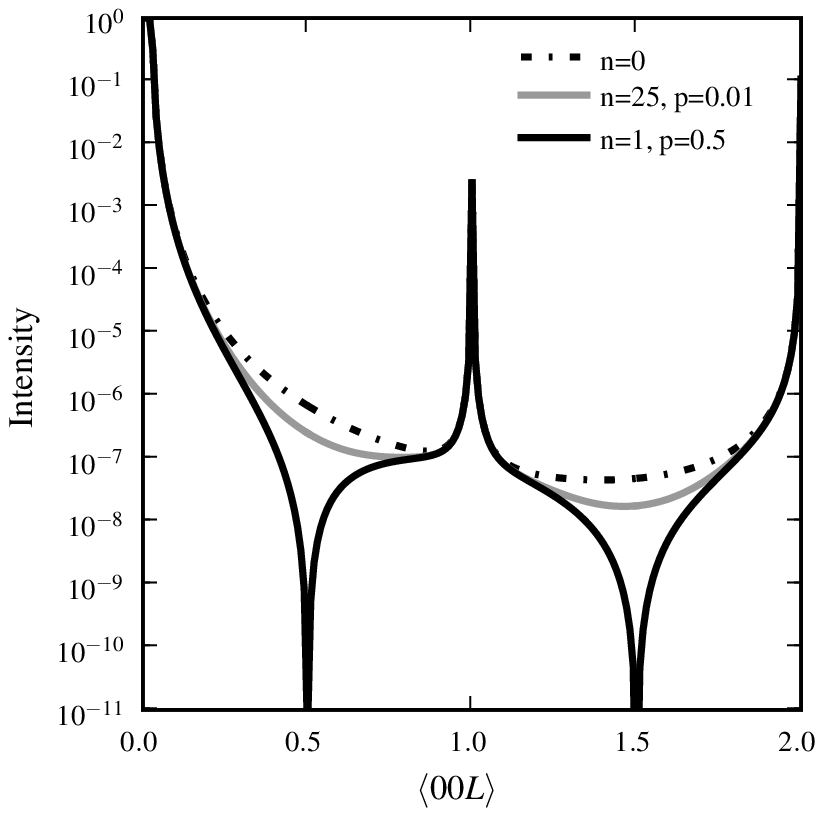}
      \caption{Darren Dale, Physical Review B}
    \label{fig:compare-p}
\end{figure}

Figure \ref{fig:compare-p}: Calculated intensity along the specular CTR. For n=0, no material has been deposited and the substrate is ideally flat. For one pulse that covers half of the flat surface with new material, complete destructive interference is observed in the anti-Bragg scattering geometries. A nearly identical RMS surface roughness is observed for 25 pulses each randomly covering 1\% of the surface with new material, but the RMS surface roughness alone is not sufficient to model the CTR intensity profile.

\begin{figure}[tb!]
    \centering
    \includegraphics[clip=true] {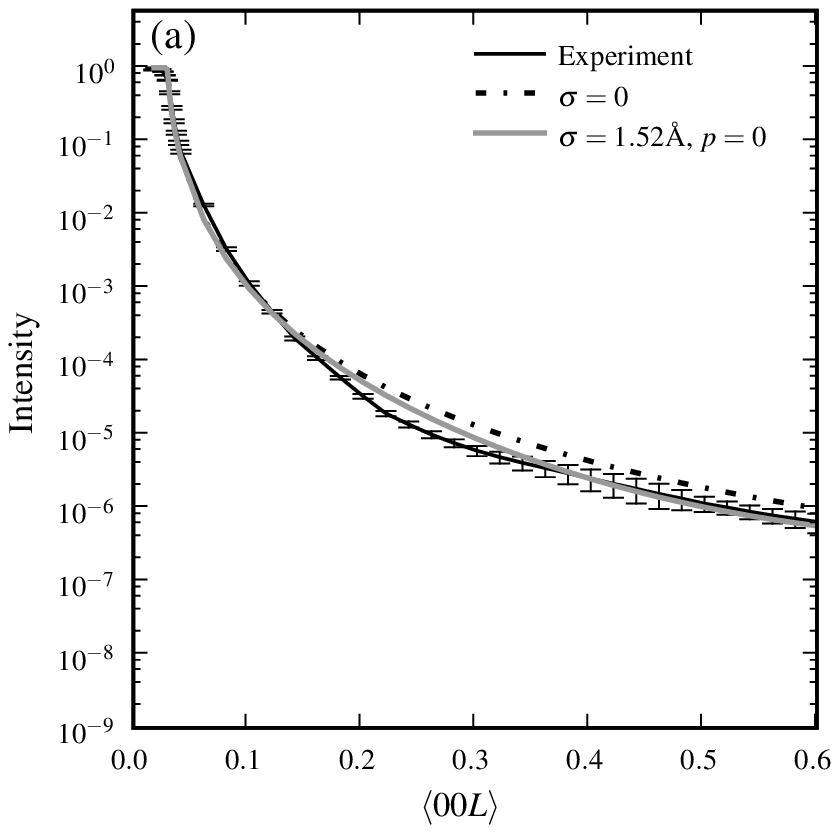}
    \includegraphics[clip=true] {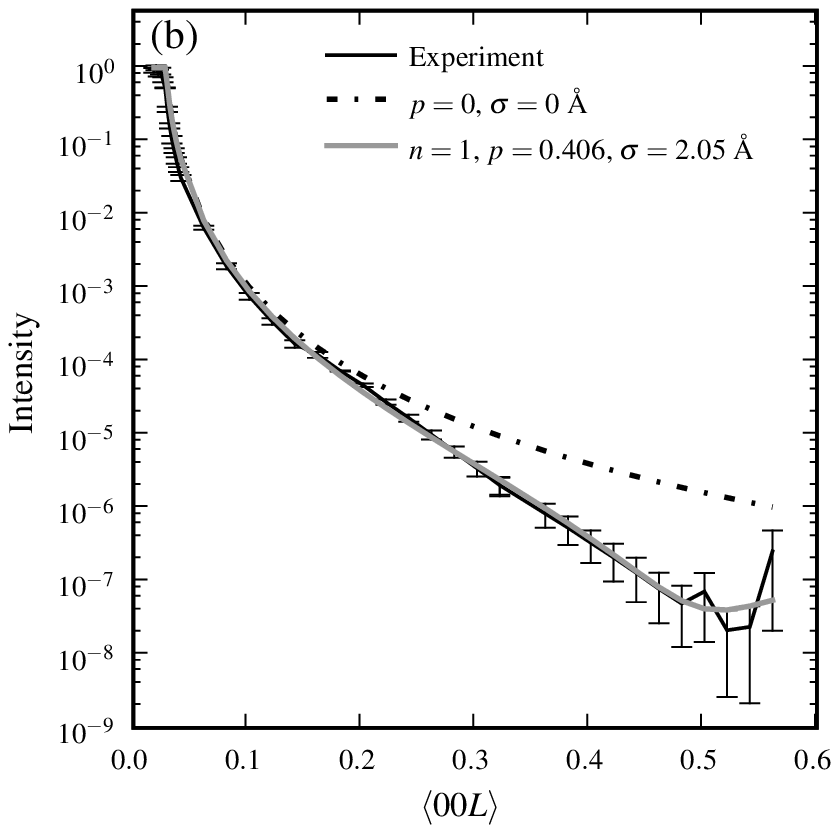}
    \caption{Darren Dale, Physical Review B}
    \label{fig:IvsL}
\end{figure}

Figure \ref{fig:IvsL}: Reflectivity data for $\langle 001 \rangle$ SrTiO$_3$ prepared with buffered-HF etch. The orientation of the miscut, and the resulting truncation rod splitting, were integrated by the resolution function. (a) 25$^\circ$C, $4\times10^{-7}$ Torr. (b) 300$^\circ$C, $1\times10^{-5}$ Torr.

\begin{figure}[tb!]
  \centering
  \includegraphics[clip=true] {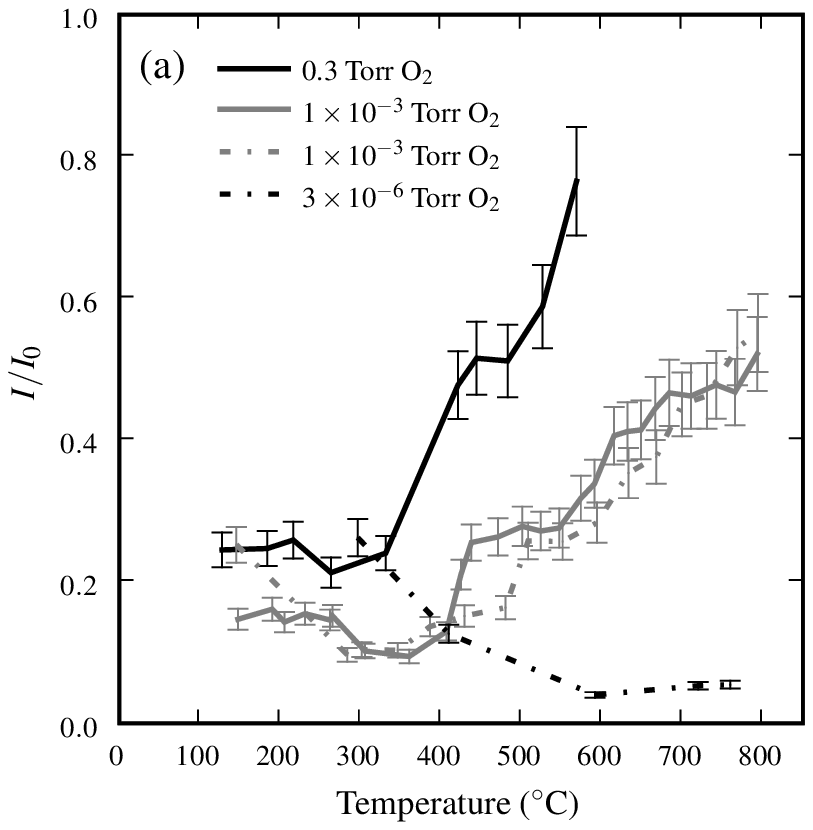}
  \includegraphics[clip=true] {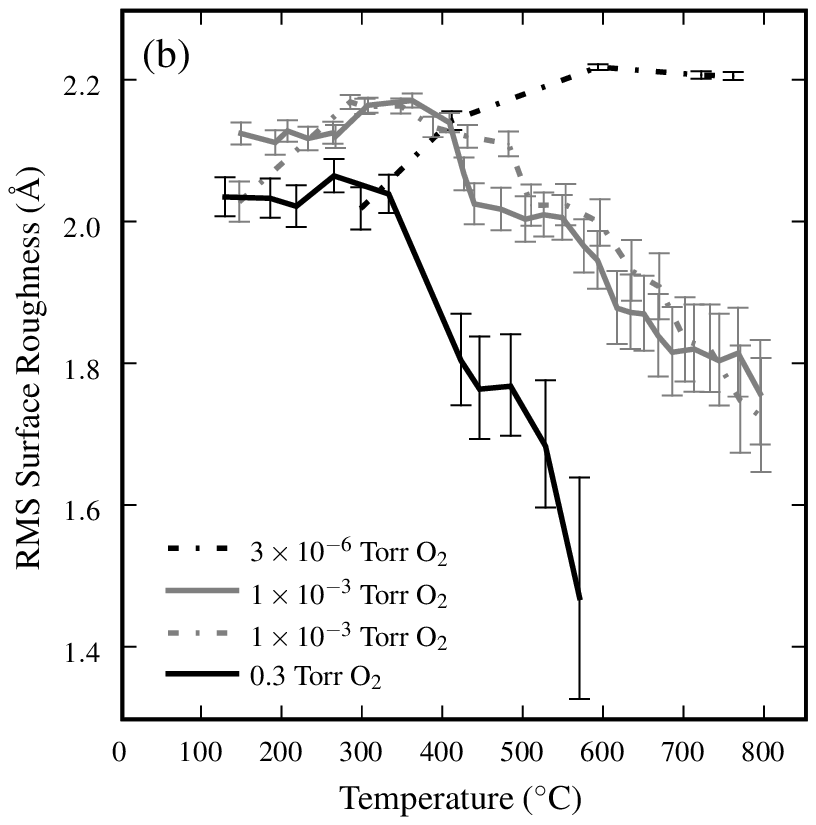}
    \caption{Darren Dale, Physical Review B}
  \label{fig:sto-roughnes-vs-T}
\end{figure}

Figure \ref{fig:sto-roughnes-vs-T}: X-ray scattering studies of \sto\ surface vs. temperature. (a) $\langle 00\frac{1}{2} \rangle$ scattering intensity vs. temperature. (b) RMS surface roughness vs. temperature, as determined from the scattering intensity.

\begin{figure}[tb!]
  \centering
  \includegraphics[clip=true] {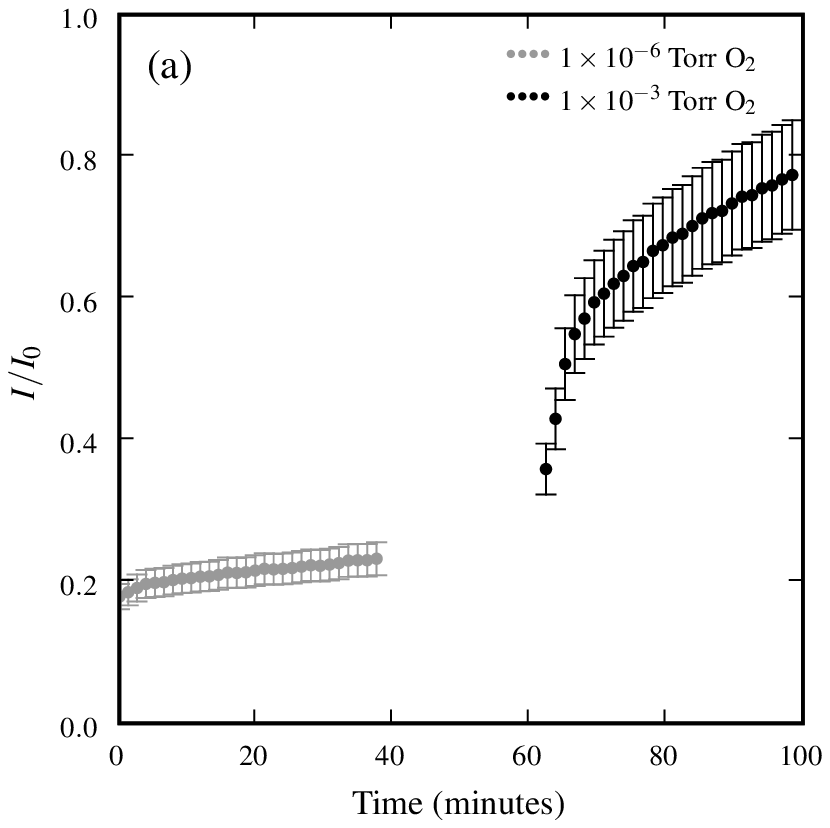}
  \includegraphics[clip=true] {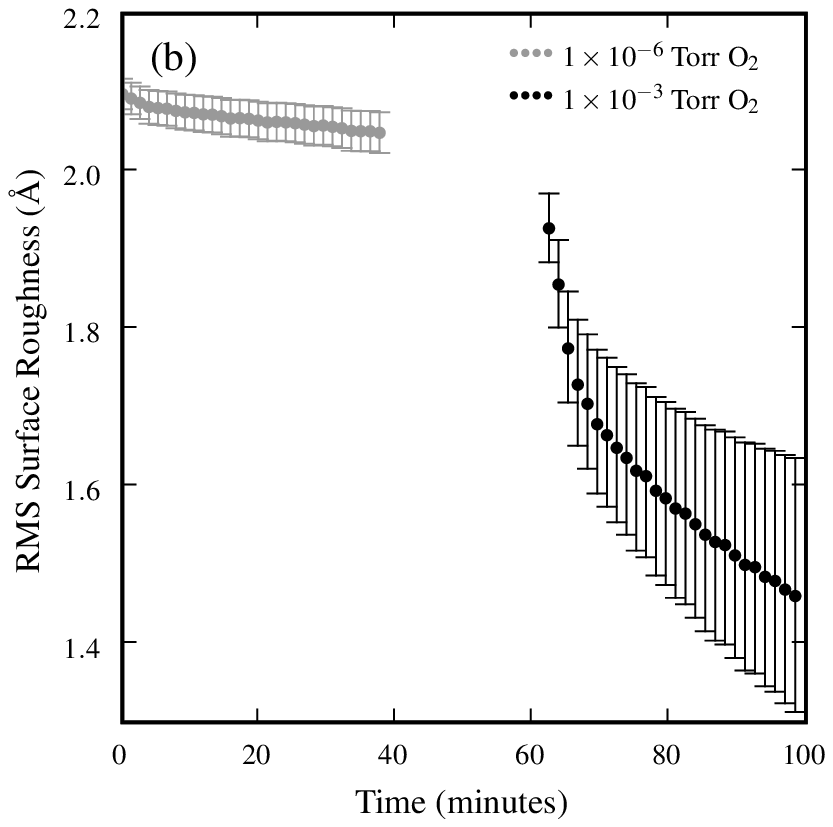}
    \caption{Darren Dale, Physical Review B}
  \label{fig:sto-roughnes-vs-time}
\end{figure}

Figure \ref{fig:sto-roughnes-vs-time}: X-ray scattering measurement of \sto\ surface vs. time. The sample temperature is 800$^\circ$C, and the O$_2$ pressure is increased from $1\times10^{-6}$ Torr to $1\times10^{-3}$ Torr after one hour. (a) $\langle 00\frac{1}{2} \rangle$ scattering intensity vs. temperature. (b) RMS surface roughness vs. temperature, as determined from the scattering intensity.

\begin{figure}[tb!]
    \centering
    \includegraphics[clip=true] {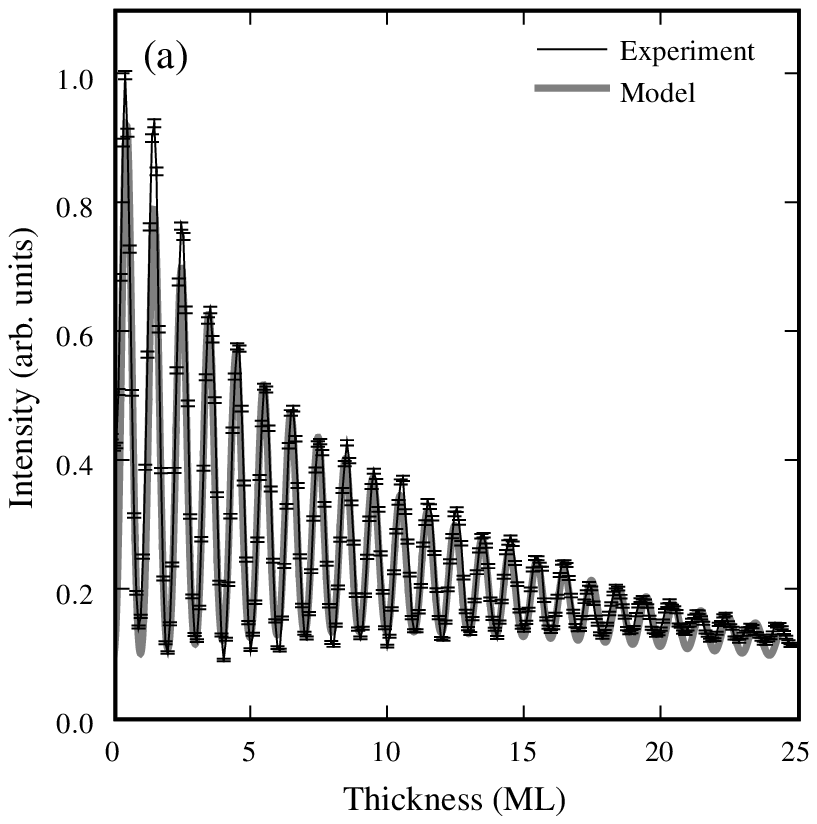}

    \includegraphics[clip=true] {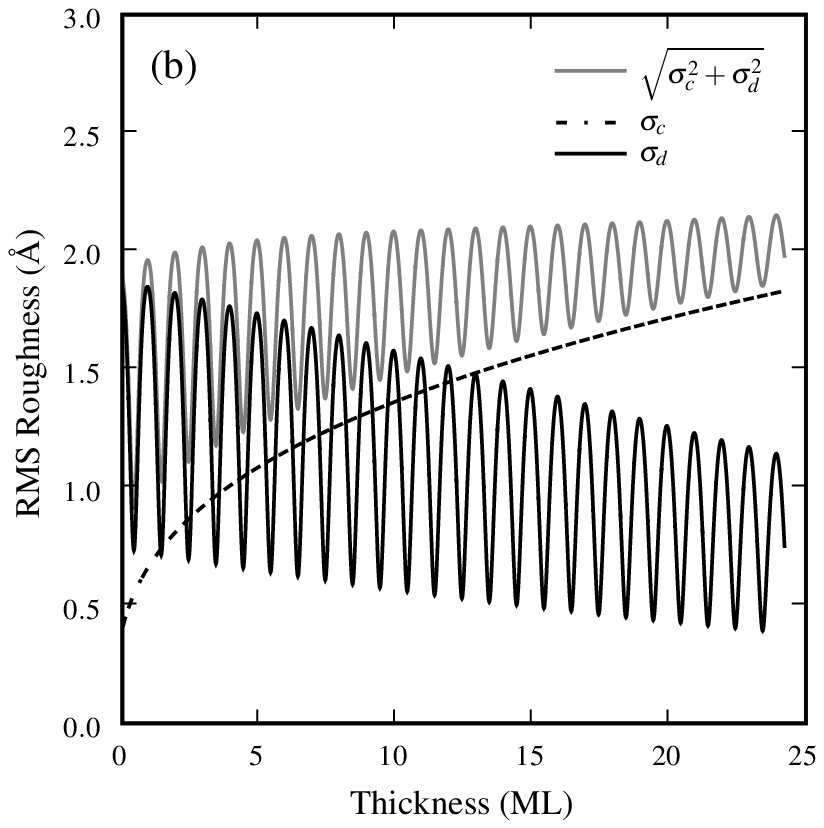}
      \caption{Darren Dale, Physical Review B}
    \label{fig:AB-oscillations}
\end{figure}

Figure \ref{fig:AB-oscillations}: Anti-Bragg X-ray scattering measured \textit{in-situ} during $\langle 001 \rangle$ \sto\ homoepitaxy via PLD. (a) Growth oscillations measured in the $\langle 00\frac{1}{2}\rangle$ scattering geometry. The model uses the roughness depicted in (b), where the total surface roughness includes both continuous and discrete contributions, which add in quadrature.

\clearpage

\clearpage

\clearpage

\clearpage

\clearpage

\clearpage

\clearpage

\end{document}